# LaTiO$_x$N$_y$ thin film model systems for photocatalytic water splitting: physicochemical evolution of the solid-liquid interface and the role of the crystallographic orientation

*Markus Pichler[a], Wenping Si[a], Fatima Haydous[a], Helena Téllez[b], John Druce[b], Emiliana Fabbri[c], Mario El Kazzi[c], Max Döbeli[d], Silviya Ninova[e], Ulrich Aschauer[e], Alexander Wokaun[a], Daniele Pergolesi[a,\*], and Thomas Lippert[a,f,g,\*]*

[a] Neutron and Muon Research Division, Paul Scherrer Institut, 5232 Villigen-PSI, Switzerland

[b] Electrochemical Energy Conversion Division, International Institute for Carbon-Neutral Energy Research (I2CNER), Kyushu University, 744 Motooka, 819-0395, Fukuoka, Japan

[c] Energy and Environment Research Division, Paul Scherrer Institut, 5232 Villigen-PSI, Switzerland

[d] Ion Beam Physics, ETH Zurich, 8093 Zurich, Switzerland

[e] Department of Chemistry and Biochemistry, University of Bern, Freiestrasse 3, 3012 Bern, Switzerland

[f] Laboratory of Inorganic Chemistry, Department of Chemistry and Applied Biosciences, ETH Zurich, 8093 Zurich, Switzerland

[g] Molecular Photoconversion Devices Division, International Institute for Carbon-Neutral Energy Research (I2CNER), Kyushu University, 744 Motooka, 819-0395, Fukuoka, Japan

\* thomas.lippert@psi.ch, daniele.pergolesi@psi.ch

**Abstract**

The size of the band gap and the energy position of the band edges make several oxynitride semiconductors promising candidates for efficient hydrogen and oxygen production under solar light illumination. The intense research efforts dedicated to oxynitride materials have unveiled the majority of their most important properties. However, two crucial aspects have received much less attention. One is the critical issue of the compositional/structural surface modifications occurring during operation and how these affect the photoelectrochemical performance. The second concerns the relation between the electrochemical response and the crystallographic surface orientation of the oxynitride semiconductor. These are indeed topics of fundamental importance since it is exactly at the surface where the visible light-driven electrochemical reaction takes place.

In contrast to conventional powder samples, thin films represent the best model system for these investigations. This study reviews current state-of-the-art of oxynitride thin film fabrication and characterization before focusing on $LaTiO_2N$ selected as representative photocatalyst. We report the investigation of the initial physicochemical evolution of the surface. Then we show that, after stabilization, the absorbed photon-to-current conversion efficiency of epitaxial thin films can differ by about 50% for different crystallographic surface orientations and be up to 5 times larger than for polycrystalline samples.

## 1. Introduction

The world's ever increasing energy demand and the enormous environmental impact of current power generation systems are among the major and most urgent problems modern societies have to face. The most obvious and intrinsically sustainable source of energy is in front of us every day, the sun that delivers a virtually unlimited amount of energy in form of light and heat. The direct conversion of the solar energy to electricity and its storage as chemical energy, e.g. as clean fuel, such as hydrogen, are certainly among the most promising strategies to become more independent from fossil and nuclear fuels, thus dramatically cutting our environmental impact on the planet. The conversion of solar light to electricity (photovoltaic) and to chemical energy by harvesting hydrogen from water (photoelectrochemical water splitting), rely both on semiconductor materials with band gaps matching the solar spectrum. The photo-generated charge carriers can be collected to directly produce an electric current or they can be used to run electrochemical reactions at the semiconductor surface to split water molecules to produce oxygen and hydrogen gas. The latter can then be used as carbon-neutral fuel for internal combustion engines or fuel cells.

This study focuses on oxynitride materials, a specific class of semiconductors that can be applied as photocatalyst in an artificial photosynthesis process that uses the photo-generated electrons and holes to split water in a photoelectrochemical cell.

The race to discover a photocatalyst that can efficiently use the sunlight to decompose water into hydrogen and oxygen[1] began about 45 year ago after the historic work of Fujishima and Honda[2] when the solar water splitting was first demonstrated using $TiO_2$.

As many other oxide materials, $TiO_2$ has a band gap well above 3 eV. The photo-response peaks in the UV energy range, thus only a few percent of the solar spectrum can be effectively used. Since many years, the main target of photocatalysis is to find materials with more suitable band gaps, enabling a more efficient use of the solar radiation. The band gap must of course be larger than 1.23 eV, which is the energy required to split the water molecule, and a value around 2 eV would make it possible to use more of the solar radiation. But this is not the only criterion a good photocatalyst must meet. Besides being chemically stable in water, a good photocatalyst must also offer good mobility to the photo-generated electrons and holes avoiding their recombination before performing the reduction and oxidation reactions at its surface. Finally the absolute position in energy of the band edges is also important; to allow the oxidation and reduction reactions (overall water splitting), the band gap should encompass both the hydrogen and the oxygen evolution potentials at 0 and 1.23 V vs. reversible hydrogen electrode (RHE).

To find a material that satisfies all these criteria is a mission Materials Science has not yet accomplished. However, many oxynitride materials are very good candidates, mainly concerning both the size and energy position of the band gap. Oxynitrides of transition metals, alkaline earth metals and rare earth metals are oxide materials where N is partially substituted into the oxygen sites. The main effect of the N substitution on the electronic (and optical) properties is the reduction of the bandgap of the pristine oxide due to the hybridization of the N 2p orbitals at higher energy levels with the O 2p orbitals[3] at the valence band maximum. This leads to the creation of additional energy levels available above the valence band maximum, thus reducing the band gap.

Over the last decade many theoretical and experimental efforts have been devoted to the development of innovative nanostructure, sample design, as well as new materials to improve the efficiency of the photoelectrochemical process.[3]

In addition, some oxynitrides have also been investigated as magnetic and dielectric materials,[4] as well as biomedical coatings[5] and pigments.[6]

Oxynitride samples are commonly proposed as powders by thermal ammonolysis of oxide precursors[7] and it was just the use of powders that allowed unveiling most of the properties that make these materials particularly promising for solar water splitting. However, specific material properties cannot be probed with powder samples. Polycrystalline powder samples do not provide for instance well-defined surfaces for a detailed investigation of the solid/liquid interface, which is where the electrochemical reaction takes place. Moreover, the diverse arrangement at atomic level of the constituent elements along surfaces with different crystallographic orientations may also play a role enhancing or inhibiting the surface electrochemistry. Such investigation has never been addressed so far for oxynitrides.

In this respect, thin films are ideal model systems to investigate surface and interface properties and have proved to be an invaluable tool for a number of different disciplines. Besides providing very well-defined surfaces, the use of thin films offer a way to tune the composition, e.g. the N content in oxynitride materials, and the micro-morphology of the sample in terms for example of average grain size. Also, the crystalline properties of the sample can be controlled growing films which are amorphous, polycrystalline, as well as highly ordered epitaxially oriented with different crystallographic surface orientations.

However, in spite of the extensive available literature on oxynitride thin films fabrication, studies using these films to address their photoelectrochemical characterization are surprisingly scarce.

## 2. Thin Oxynitrides Film as Model System in Photoelectrochemical Water Splitting

### 2.1. Deposition Techniques of Oxynitride Thin Films

Oxynitride thin films can be grown by chemical vapor deposition (CVD), including atomic layer deposition (ALD), and physical vapor deposition (PVD) including reactive magnetron sputtering and pulsed laser deposition (PLD).

#### 2.1.1. Chemical Vapor Deposition

Amorphous $TiO_xN_y$ film has been deposited by CVD using $NH_3$ plasma at a deposition temperature of at 300 ºC and different titanium precursors, such as tetraisopropoxide (TTIP) and tetrakis(diethylamido)titanium (TDEAT). The latter allowed the largest nitrogen incorporation of 28%.[8]

$SiO_xN_y$ on-chip interconnects for nano optical/photonics applications were developed by plasma enhanced CVD. 80 nm thick films of $SiO_xN_y$ were grown on Si wafers at 300 ºC using $SiH_4$, $N_2O$, $NH_3$ as precursors.[9] Details on the film composition were not provided. A post annealing in vacuum at 600 ºC was required to achieve the desired morphological and optical properties by improving the film densification and reducing the pronounced surface roughness of the as-grown films.

$TiO_xN_y$ films were prepared by plasma enhanced ALD using tetrakis(dimethylamino)titanium (TDMAT) with $N_2$ plasma or TTIP with a $NH_3$ plasma in the temperature range of 170-450 ºC.[10] Higher nitrogen contents in the films were obtained by using the nitrogen-rich TDMAT precursor and $N_2$ plasma than by using oxygen-rich TTIP and $NH_3$ plasma.[10a] The O/N ratio in the $TiO_xN_y$ films was found to be critical for the electric properties such as conductivity, dielectric breakdown, and permittivity.[10b]

ALD combined with periodic oxidation was used to prepare $TiO_xN_y$ films wherein TiN was grown from titanium tetrachloride and $NH_3$ and oxygen was supplied periodically between the ALD cycles.[11] The periodic oxidation enabled tuning the nitrogen contents in the films, which ranged between 10% and 30%.[11]

Lithium phosphorus oxynitride (LiPON) films used as solid-state electrolyte for Li-ion rechargeable batteries were developed *via* ALD using different phosphorus and lithium precursors with nitrogen as carrier/purge gas.[12] Relatively small room temperature ionic conductivity ranging from 1 to $7 \times 10^{-7}$ S cm$^{-1}$ was obtained depending on the N content. LiPON thin films were actually for the first time fabricated by PVD, specifically by sputtering a $Li_3PO_4$ target in pure $N_2$ in 1993.[13] Nine years later also PLD was effectively applied for the fabrication of LiPON films by ablating a $Li_3PO_4$ target in a $N_2$ gaseous environment.[14] In both cases, Li-ion conductivity around $2 \times 10^{-6}$ S cm$^{-1}$ at room temperature was achieved.

In general, literature reports that the most critical issue to be addressed for the growth of oxynitride films by CVD is related to the complex chemical reactions involved in the process. To obtain the expected compositions, the precursor/co-reactant combination must be carefully selected since the oxygen content in the precursor and oxygen residuals in the reactor strongly affect the O/N ratio in the film.[8, 10a]

To our knowledge, no oxynitride photocatalysts fabricated as thin films by CVD have been tested toward their PEC performance.

*2.1.2. Physical Vapor Deposition*

Reactive magnetron sputtering is the most commonly used PVD technique for preparing oxynitrides films. In a typical deposition process, reactive gas mixtures, such as $O_2$-$N_2$,[15] Ar-$N_2$,[16] Ar-$N_2$-$O_2$[17] as well as Ar-$N_2O$[18] are used. The sputtering targets can be either the oxynitride,[19] the related oxide[20] or (for ternary oxynitrides) a single element target.[18, 21] Among the oxynitride thin film materials grown by reactive magnetron sputtering are LaTiO$_x$N$_y$,[19, 22] TiO$_x$N$_y$,[17, 23] Ti(Cr)O$_2$:N,[24] TaO$_x$N$_y$,[15, 21] Zn$_x$O$_y$N$_z$,[16] SiO$_x$N$_y$,[18] (Sr$_{0.99}$La$_{0.01}$)(Ta$_{0.99}$Ti$_{0.01}$)O$_2$N,[20] V$_x$ON and Cr$_x$ON.[25] The obtained compositions bang gap energies and crystalline properties of the films can be adjusted by tuning the sputtering parameters, such as the substrate temperature and the $N_2$ content of the reactive gas mixture.[15, 19] Relatively low substrate temperatures were used for the deposition of ternary oxynitride films, for example, TiO$_x$N$_y$ (250-400 °C),[17, 23a] TaO$_x$N$_y$ (room temperature-100 °C),[15, 21] Zn$_x$O$_y$N$_z$ (150 °C),[16] and SiO$_x$N$_y$ (350 °C),[18] while higher temperatures were needed for quaternary and more complex oxynitride films, for example, LaTiO$_x$N$_y$ (800-900 °C),[19, 22] and $(Sr_{0.99}La_{0.01})(Ta_{0.99}Ti_{0.01})O_2N$ (750 °C).[20] Polycrystalline, oriented, and epitaxial films with different N contents and bandgap energies have been reported.[16-21]

PLD is an effective method to prepare high quality oxynitride thin films. Heteroepitaxial $CaTaO_2N$ films have been deposited on (100)-oriented $SrTiO_3$ substrate by nitrogen plasma-assisted PLD with a $Ca_2Ta_2O_7$ target at a substrate temperature of 800 °C under a $N_2$ partial pressure of 100 mTorr.[26] Epitaxial $BaTaON_2$ films were deposited on $SrRuO_3$-buffered $SrTiO_3$ (100) using a $BaTaON_2$ target at a substrate temperature of 760 °C in a 20:1 ratio $N_2/O_2$ gas mixture at a pressure of 100 mTorr.[4b]

Recently, pulsed reactive crossed beam laser ablation (PRCLA) was used for the growth of $LaTiO_xN_y$ thin films. PRCLA is a modification of standard PLD where, instead of setting a uniform background pressure of the gaseous environment, pulsed gas jets of a reactive gas (typically $NH_3$ is used for the growth of oxynitride films) are injected near the ablation spot at the target surface.[27] Several different substrates were used, also conducting substrate for the fabrication of complete photoanodes for PEC measurements. As for sputtering and PLD, also for PRCLA different deposition parameters allow the growth of polycrystalline, oriented, and epitaxial $LaTiO_xN_y$ films with different N content.[28]

**2.2. Photoelectrochemical Measurements of Oxynitride Thin Films**

It is important to highlight first that, as for powder samples, also for thin films the PEC performance reported in literature are often difficult to compare, even for the same oxynitride material. On one side the fabrication method and sample design, including the selection of the current collector, significantly affect the PEC response. On the other side, the PEC tests are performed in different experimental conditions: using different electrolytes compositions and pH values, using $[Fe(CN)_6]^{3-}/[Fe(CN)_6]^{4-}$ redox couples as sacrificial reagents, with or without the use of different kind of cocatalysts. Furthermore, the initial physicochemical evolution of the surface in operating conditions leads to relatively stable PEC response after a number of transient states which depends on the experimental parameters (electrolyte composition, voltage scan, holding potential time at the highest applied voltage). Typically the stabilized photocurrent density is significantly smaller than that measured during the initial tests but unfortunately whether the reported results actually refer to stabilized conditions is rarely specified.

As a final general observation, the photocurrent densities measured using model system thin films are expected to be significantly lower than the values achievable with powder samples. As will be discussed next, while for thin films the illuminated area and the area of the photoanode exposed to the electrolyte (thus potentially electrochemically active) coincide, for powder samples the latter can be up to 50 times larger than the illuminated area.

After these premises, we move forward noticing that only few studies report the investigation of the PEC performance of oxynitride thin films towards solar water splitting.

The PEC behavior of $TaO_xN_y$ photoanodes was investigated by Leroy *et al*.[21] The films were deposited at room temperature on fluorine-doped tin oxide substrates by DC reactive sputtering using a Ta target under a mixture of Ar, $O_2$, $N_2$ gases. The chemical composition of the films was investigated by Auger electron spectroscopy. The PEC characterization, performed in a neutral $Na_2SO_4$ electrolyte, showed that the largest photocurrent density of 30 $\mu A\ cm^{-2}$ at 0.7 V vs. RHE was achieved for films grown in an oxygen partial pressure of $1.9 \times 10^{-4}$ mbar during the sputtering process. This value of oxygen partial pressure resulted in an oxygen content in the film leading to a good balance between the chemical stability, which is poor for low oxygen contents, and the width of the band gap, which become too large when the oxygen content increases too much.

LaTiO$_x$N$_y$ (LTON) thin films deposited by reactive radio-frequency sputtering on electronically conducting Nb-doped strontium titanate substrates were studied as photoelectrodes for water splitting.[19] The deposited thin films were (001)-epitaxially oriented, oriented (not epitaxial), and polycrystalline depending on the substrate temperature and plasma composition. The chemical composition of the films was investigated by energy-dispersive X-ray spectroscopy. The films with higher N content were polycrystalline, while those with lower N content were the epitaxially oriented. The photoelectrochemical measurements performed in [Fe(CN)$_6$]$^{3-}$/[Fe(CN)$_6$]$^{4-}$ electrolyte of pH 7.8 showed a photocurrent density of 14 µA cm$^{-2}$ at 0.2 V vs. Ag/AgCl for the epitaxial films. This value is about a factor of 2 higher than those measured with the oriented and polycrystalline films. It was concluded that the positive effect of higher crystallographic quality that reduces the electron/hole recombination in the epitaxial films dominates over the negative effect of the lower N content that limits the photo-response in the visible light energy range. The photoactivity of the epitaxial LTON film was further enhanced by modification of the surface with a IrO$_x$ cocatalyst which led to a 4 fold increase in the photocurrent density at +0.5V vs. Ag/AgCl in an aqueous Na$_2$SO$_4$ solution of pH=4.5.[19]

The PEC characterization of LTON thin films is also reported for polycrystalline samples grown by PRCLA.[28] In that study, the photoanode consisted of the LTON semiconductor grown on MgO single crystal substrates coated by DC reactive sputtering with a buffer layer of TiN as a current collector. The cation composition of the LTON films was measured by Rutherford Backscattering (RBS) and the N/O ratio was determined by Elastic Recoil Detection Analysis (ERDA). Using sodium borate electrolyte buffered at pH of 9, without any cocatalysts or sacrificial reagents, a photocurrent density of about 0.3 µA cm$^{-2}$ at about 1.25 V vs. RHE was measured. The photocurrent density was probably limited by the extended grain boundary region along the LTON layer that facilitates the recombination of the photo-generated charge carriers. Moreover, the TiN-coated substrates were fabricated ex situ; the sample transfer and manipulation, may have favored the formation of an interfacial layer that further enhanced the charge recombination.

However, the use of TiN-buffered substrates showed potential for tuning the N content and the crystallographic properties of LTON films grown by PRCLA. The availability of films with different micro-morphologies and different crystallographic surface orientations would allow to distinguish not only the role of a different crystalline quality on the PEC performance but also for the first time the role of the crystallographic orientation of the surface, where the visible light-driven electrochemical reaction takes place.

Here, we report further advances on the development of model LTON photoanodes grown as thin films on TiN current collectors. Polycrystalline LTON films are first used to probe the physicochemical modification of the surface induced by the photoelectrochemical tests. Finally, the use of TiN layers grown with different crystallographic orientations on different substrates allows the comparison of the photoelectrochemical performance of polycrystalline, (112)-oriented (not epitaxial), as well as (001) and (011) epitaxially oriented LTON thin films.

## 3. Results and Discussion

For this study $LaTiO_xN_y$ thin films were grown by PRCLA on (001)-oriented MgO and (0001)-oriented $Al_2O_3$ single crystal substrates coated with TiN layers epitaxially grown in situ by conventional PLD and used as current collector for PEC tests. Details on the PRCLA method can be found in ref. [27-29], as well as in the experimental section. As expected, and as observed in our previous study, the TiN layer does not give any contribution to the measured photo-current (see Supporting Information, Figure S1).

## 3.1. Structural, Morphological, Chemical, and Optical Characterizations

**Figure 1** illustrates the X-Ray Diffraction (XRD) measurements of the four samples. The samples include one polycrystalline (LTON-poly) and three oriented LaTiO$_x$N$_y$ thin films grown on TiN-buffered substrates. The three oriented films are labeled LTON-hkl, where hkl indicates the preferential orientation.

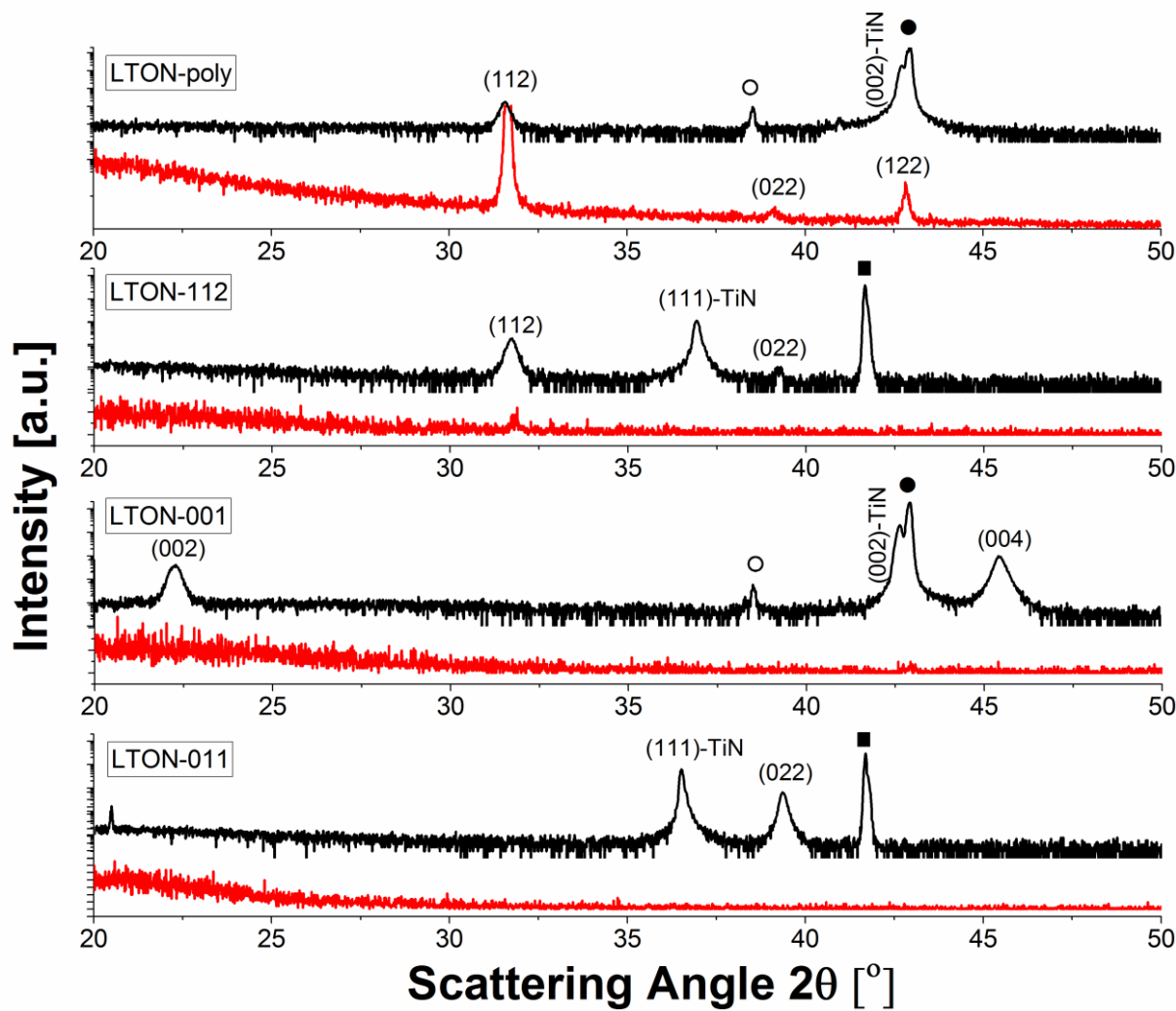

**Figure 1.** XRD pattern of LTON-poly, LTON-112, LTON-001 and LTON-011 shown by (black) the θ/2θ-scan and (red) in the grazing incidence mode. The Al$_2$O$_3$ substrate is marked with a ■, the MgO substrate is marked with a ● and the K$_\beta$ diffraction reflex from the MgO substrate is marked with a ○.

LTON-poly and LTON-001 were grown on TiN-buffered (001)-oriented MgO while LTON-112 and LTON-011 were grown on TiN-buffered (0001)-oriented Al$_2$O$_3$. The lattice parameters and crystal structures of MgO, $Al_2O_3$, TiN and $LaTiO_2N$ are reported in **Table 1**. The TiN buffer layer was grown epitaxially on both substrates and is used in this study as current collector for photoelectochemical characterization of the oxynitride layers. TiN grows cube-on-cube epitaxially-oriented on MgO[30] due to the relatively small lattice mismatch of about 0.56%. The epitaxial relation between the (111)-oriented TiN surface and (0001)-oriented $Al_2O_3$ surface is discussed in detail in ref. [31]. It is shown that the very large lattice mismatch of 8.46% is accommodated at the interface by a periodic arrangement of misfit dislocations leading to a highly ordered fully relaxed growth of epitaxial films showing a FWHM of the rocking curve measured at the (111) TiN reflex as low as 0.07°.

**Table 1.** Lattice parameters of the two substrates MgO (ICSD Coll.Code: 158103) and $Al_2O_3$ (ICSD Coll.Code: 160604), the TiN buffer layer (ICSD Coll.Code: 152807) and the oxynitride photocatalyst $LaTiO_2N$ (ICSD Coll.Code: 168551).

| Compound | Crystal structure | Lattice parameters | | |
|---|---|---|---|---|
| | | a | b | c |
| MgO | fcc | 4.21130 | | |
| $Al_2O_3$ | hexagonal | 4.76170 | | 12.99900 |
| TiN | fcc | 4.23500 | | |
| $LaTiO_2N$ | orthorhombic perovskite | 5.60279 | 5.57137 | 7.87900 |

The θ/2θ-scan of LTON-poly mainly shows the (112) reflex (Figure 1), which is the crystallographic orientation with the largest relative intensity for this material. The GI-XRD (grazing incidence) measurements reveal a strong peak of the same reflex, but also the presence of the (022) and (122) reflexes confirming the polycrystalline structure of this sample.

LTON-112 shows a strong (112) reflection peak and traces of a minor (022) orientation. However, the GI-XRD measurement only shows one and very weak reflex at the (112) position (Figure 1). This suggests a highly textured microstructure of this film, characterized by adjacent grains (112)-oriented parallel to the substrate surface normal.

LTON-011 and LTON-001 grew epitaxially on (111)-oriented TiN-buffered $Al_2O_3$ and (001)-oriented MgO substrates, respectively. Neither sample shows any reflexes in the GI-XRD measurements (Figure 1) confirming the epitaxial growth.

Our previous study reports that the crystallographic properties and nitrogen content of $LaTiO_xN_y$ grown on (001)-oriented TiN-buffered MgO substrate depend on the laser fluence.[28] High laser fluences ensure high nitrogen contents, but also lead to a polycrystalline $LaTiO_xN_y$ thin film. Lowering the laser fluence allows the eptiaxial growth of $LaTiO_xN_y$, but at the same time decreases the nitrogen content. The same effect of an improved crystalline quality with decreasing the nitrogen content has been reported also for $LaTiO_xN_y$ films grown by sputtering.[19] For this work, TiN-coated MgO and $Al_2O_3$ substrates were used to grow $LaTiO_xN_y$ films at high laser fluences (above 3 J cm$^{-2}$) and at lower laser fluences (about 2 J cm$^{-2}$) keeping all other deposition parameters constant.

The result obtained on (001)-oriented TiN-coated MgO confirms our previous finding concerning the crystallographic properties of the oxynitride films. In particular, we assume that in spite of the large lattice mismatch of about 6.6% between TiN and $LaTiO_xN_y$ the epitaxial growth can proceed assisted by the introduction of interfacial misfit dislocations that fully release the excess stress.[28]

Concerning the films grown on TiN-coated sapphire substrates, the epitaxial growth of (011)-oriented $LaTiO_xN_y$ films on (111)-oriented TiN surfaces can be rationalized by assuming a domain matching epitaxy where 2 unit cells of $LaTiO_xN_y$ match with 1 unit cell of TiN with a lattice misfit of about 6.5% accommodated by a large density of interfacial misfit dislocations. On the contrary, no lattice matching between the (112)-oriented $LaTiO_xN_y$ surface and the (111)-oriented TiN could be identified. We assume that the driving force for the (112)-oriented growth of a textured film is in this case a combination between the selected deposition parameters and the low Gibbs' free energy of this surface. As previously mentioned, the (112) XRD reflex shows in fact the highest relative intensity for this material.

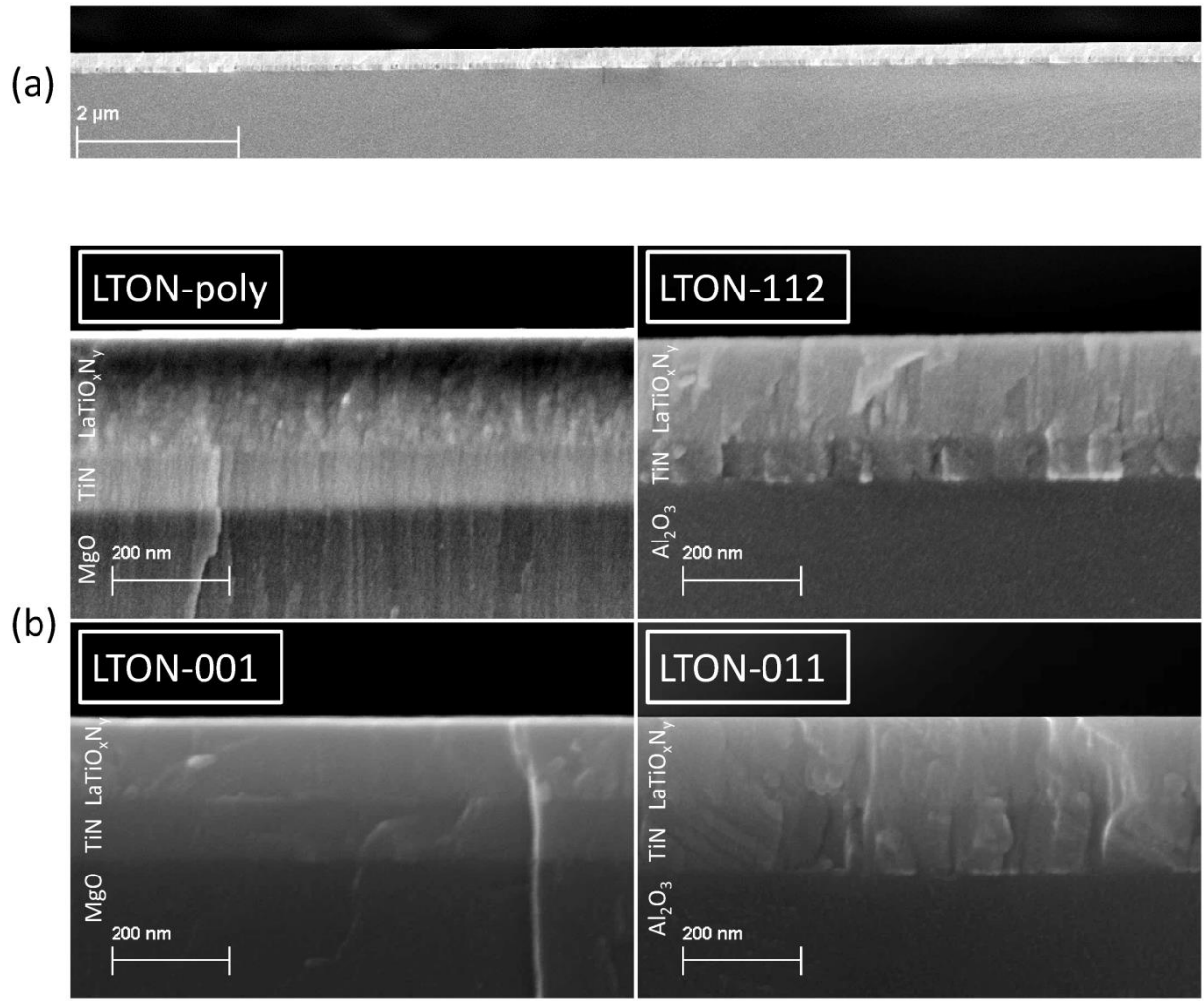

**Figure 2.** SEM cross sections (fractured surface) of LTON-poly, LTON-112, LTON-001 and LTON-011.

Scanning Electron Microscopy (SEM) was used for morphological characterization. **Figure 2**a shows a representative example of a cross section SEM micrograph of one of the samples fabricated for this work. All films reveal good thickness uniformity. No significant evidence of microscopic defects, such as isolated micro particles, was observed.

Figure 2b shows the acquired micrographs at higher magnification of the fractured cross sections of the four samples. The SEM cross section of LTON-poly shows a granular morphology without clear evidence of definite crystallographic structural planes. Such morphological features can be ascribed to a polycrystalline film, as revealed by XRD analysis. Moreover, no clear cleavage fractures that range through the whole sample are observed suggesting a film growth crystallographically decoupled from the TiN-coated MgO substrate.

The SEM micrographs of the other three samples show the typical features of a textured or highly ordered crystalline structure. In particular, in the epitaxial films continuous cleavage fractures are visible extending from the substrate through the TiN/LaTiO$_x$N$_y$ bi-layer. The films grown on sapphire show features ascribed to grain boundaries. Grain boundaries and cleavage planes extend over the complete bi-layer in the LTON-011 film, while they look discontinuous at the TiN/LaTiO$_x$N$_y$ interface of the LTON-112 film. This observation agrees with the previous discussion of the XRD results suggesting an epitaxial growth of LTON-011 and a highly textured but not epitaxial growth of LTON-112. LTON-001 shows a very homogeneous cross section, almost featureless with the exception of continuous cleavage fractures. It is worth mentioning here that due to the different crystalline structures of the two substrates the fractured cross section of the samples grown on MgO look in general much more regular and smooth.

Atomic-Force Microscopy (AFM) was performed to further investigate the surface topography. For the oriented samples, AFM analysis shows a very smooth surface with an RMS value of the height distribution of less than 1 nm. The polycrystalline sample has a slightly rougher surface with an RMS value of 1.5 nm.

The chemical composition of the four samples described above was investigated by RBS complemented by ERDA used to determine the nitrogen-to-oxygen ratio (**Table 2**). We have evidence of a lanthanum content that is slightly larger than the stoichiometric composition at the expense of the titanium content. This over-stoichiometric composition of lanthanum, the heaviest element in the compound, can be explained as a result of the PLD process, where the chemical content of the film of lighter elements are prone to get reduced during deposition.[32]

The thickness of the oxynitride thin films were estimated by SEM and RBS. The two measurements are in good agreement (Table 2).

**Table 2.** Composition, nitrogen-to-oxygen ratio, thickness and used laser fluence of the described samples.

| Sample | Substrate | composition by RBS | N/O-ratio by ERDA | thickness RBS | thickness SEM | Laser fluence |
|---|---|---|---|---|---|---|
| LTON-poly | TiN/MgO | $La_{1.03}Ti_{0.97}O_{2.20}N_{0.87}$ | 0.40 | 175 | 175 | 3.8 J cm$^{-2}$ |
| LTON-112 | TiN/Al$_2$O$_3$ | $La_{1.03}Ti_{0.97}O_{2.20}N_{0.90}$ | 0.41 | 155 | 170 | 3.1 J cm$^{-2}$ |
| LTON-001 | TiN/MgO | $La_{1.01}Ti_{0.99}O_{2.85}N_{0.38}$ | 0.13 | 140 | 140 | 2.2 J cm$^{-2}$ |
| LTON-011 | TiN/Al$_2$O$_3$ | $La_{1.04}Ti_{0.96}O_{2.75}N_{0.53}$ | 0.19 | 140 | 140 | 2.2 J cm$^{-2}$ |

To investigate the optical properties of our films, UV-Vis measurements were performed. **Figure 3**a shows the UV-Vis spectra acquired for three different LaTiO$_x$N$_y$ epitaxial films, about 150 nm-thick, grown on double-side polished (110)-oriented YAlO$_3$ substrates with nitrogen contents comparable to that of the samples grown on TiN-coated substrates from Table 2. The LaTiO$_x$N$_y$ samples grown on the transparent YAlO$_3$ substrates were prepared using different laser fluences following the same procedure used for the LaTiO$_x$N$_y$ samples described above. Figure 3b shows the XRD analysis of one of these samples. As shown in previous studies,[27a, 28] the reduction in laser fluence leads to a reduction of the nitrogen content and therefore also an increase in the band gap, as revealed by the shift of the absorption edge towards smaller wavelengths.

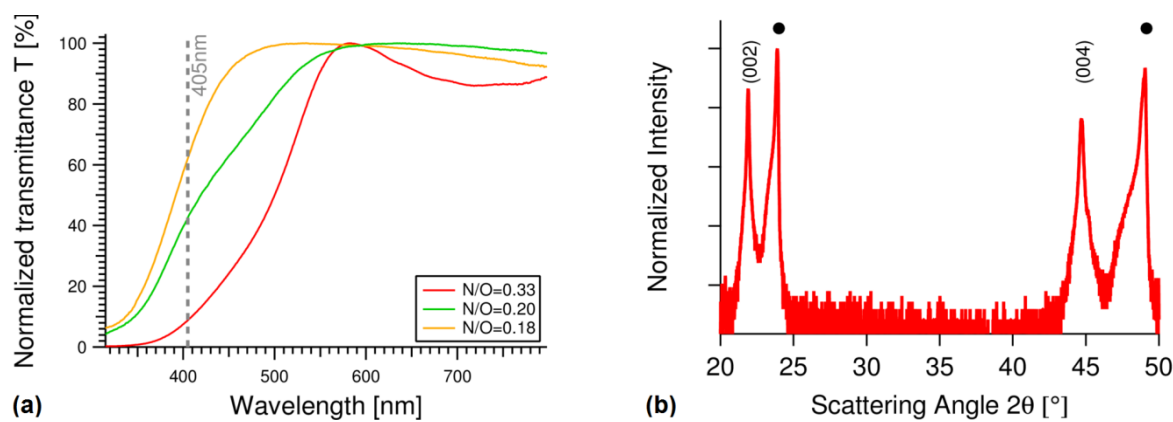

**Figure 3.** (a) Normalized UV-Vis spectra of three LaTiO$_x$N$_y$ samples grown on double-side polished YAlO$_3$ substrates with different nitrogen content to visualize the shift of the band gap to higher energies with decreasing of the nitrogen content. (b) X-ray diffraction pattern of one representative LaTiO$_x$N$_y$ sample grown epitaxially and (001)-oriented on (110)-oriented YAlO$_3$ substrate. The YAlO$_3$ substrate is marked with a ●.

Nitrogen-to-oxygen ratios of 0.33, 0.20 and 0.18 were measured by ERDA for the three films. The dashed line in Figure 3a indicates the wavelength (405 nm) of the monochromatic light source used in this study for photoelectrochemical characterization. At this wavelength, a light absorption of 91%, 57% and 38% can be estimated (neglecting reflectance) for the three samples from higher to lower nitrogen-to-oxygen ratios.

### 3.2. Photoelectrochemical Characterizations

The commonly used method to report PEC activity are potentiodynamic (PD) linear voltage scans using chopped illumination. In a PD scan the current is measured while sweeping the applied potential and turning on/off the light source.

**Figure 4**a-c show the 1$^{st}$, 2$^{nd}$ and 4$^{th}$ PD voltage scans of a LTON-poly sample. The measurements were performed applying a voltage scan rate of 10 mV s$^{-1}$, in NaOH electrolyte (0.5M) with a pH of 13 without any cocatalyst. A current density of more than 30 µA cm$^{-2}$ at 1.5 V vs. RHE was measured during the first measurement (Figure 4a). For this study the light source is a 405 nm laser diode with 5 mW power output focused on a spot area of about 0.03

cm$^2$ at the film surface. The current density is calculated considering the area of the laser spot, while the dark current which originates from the whole sample area dipped into the electrolyte was subtracted (see Supporting Information Figure S2). The surface area exposed to the electrolyte equals the area of the laser spot due to the very smooth surfaces of the samples, as observed by AFM, while for powder samples the exposed area can be about 50 times larger than the illuminated area. The typical loading density for powder samples ranges from 0.40-0.45 mg cm$^{-2}$.[7c, 33] Assuming a typically surface area measured with the BET method of about 8.5 m$^2$ g$^{-1}$ and 15 m$^2$ g$^{-1}$ [7a, 34] for LaTiO$_2$N, it can be estimated that for powder samples the area exposed to the electrolyte is about 35 to 60 times larger than the illuminated area. Considering this, the measured photocurrent density in the range of few tens of μA cm$^{-2}$ are in line with those reported in literature for LaTiO$_2$N powder samples[35] as well as for thin films,[19] where no co-catalysts or redox couples acting as sacrificial reagents were used.

The measured current density decreases, especially at higher applied potentials, with increasing number of voltage scans (Figure 4c) before stabilizing. A current density of about 10 μA cm$^{-2}$ was measured at 1.5 V vs. RHE after stabilization.

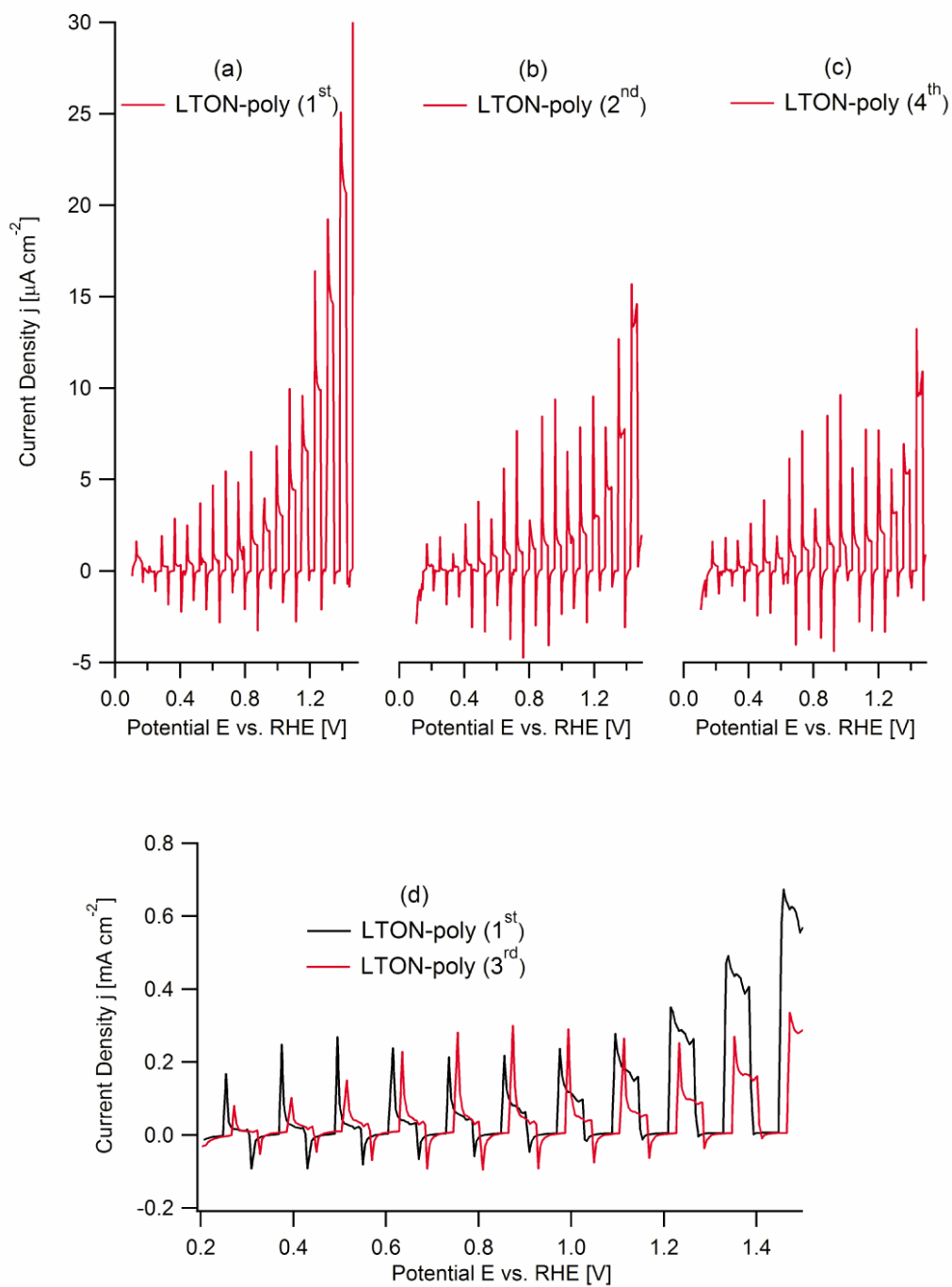

**Figure 4.** Evolution of the photocurrent: (a) 1st, (b) 2nd and (c) 4th chopped potentiodynamic measurements with 10 mV s$^{-1}$ of sample LTON-poly (4 s light on and 4 s light off) using a 405 nm laser diode with a light intensity of 130 mW cm$^{-2}$. For the ease of comparison, the dark current is subtracted from the data. (d) 1st and 3rd chopped potentiodynamic measurements with 10 mV s$^{-1}$ of LTON powder photoanode electrodeposited on fluorine-doped tin oxide substrates, measured by a 150 W Xe arc lamp with an AM 1.5G filter with an output intensity of 100 mW cm$^{-2}$. All measurements were performed in an electrolyte of 0.5 M NaOH (pH= 13).

This initial reduction of the photocurrent density is not an effect only related to thin films. Figure 4d shows the comparison between the 1$^{st}$ and the 3$^{rd}$ PD scan measured with LTON powders. The powders were produced by thermal ammonolysis and electrodeposited on fluorine-doped tin oxide substrates. A necking treatment was applied to improve the electrical contact among LTON grains and between the photocatalyst and the current collector. The PEC measurements were performed without any cocatalyst using a Xenon lamp as light source. The same electrolyte, potential range and voltage scan rate used for the characterization of the films were applied. The initial photocurrent density was about 300 µA cm$^{-2}$ at 1.23 V vs. RHE and about a factor 2.5 lower in the 3$^{rd}$ PD scan. The photocurrent density reached a constant value of about 60 µA cm$^{-2}$ after the 6$^{th}$ PD scan. At the same applied potential also the stabilized photocurrent density measured for LTON-poly (few mA cm$^{-2}$) is in line with values reported in literature for powder samples by considering the 35 to 60 times smaller area in contact with the electrolyte for thin films as compared to powders of the same material.

This reduction of the measured photocurrent within the first photoelectrochemical cycles (about 4 - 6 voltage scans) is the symptom of an initial physicochemical modification of the oxynitride semiconductor, and stability issues are well known for this class of materials.[7a, 7c, 36] In previous studies for example, nitrogen loss at the semiconductor/electrolyte interface was indicated as the primary degradation mechanism for LaTiO$_x$N$_y$ powder samples during O$_2$ evolution over a wide range of experimental conditions. [7a, 37]

### 3.3. Analysis of the PEC-Induced Surface Modifications

For this work, to further investigate the reason of the initial degradation of the PEC performance XRD, ERDA, Low-Energy Ion Scattering spectroscopy (LEIS), and Time-of-Flight Secondary Ion Mass Spectrometry (ToF-SIMS) were applied.

For these investigations, two LaTiO$_x$N$_y$ samples grown on TiN-buffered MgO substrates were prepared simultaneously (during the same deposition process). One of these samples was used for PEC characterization. Hereinafter, we refer to this sample as "after PEC", while the other non-tested sample (the LaTiO$_x$N$_y$ thin film as grown) is called "before PEC".

The samples denominated "after PEC" underwent several voltage scans up to at least 1.5 V vs. RHE and showed stable (fully stabilized) photocurrents. In our experimental setup, the area of the substrate into the electrolyte is about 5×5 mm$^2$. At the center of this area the light source used for PEC measurements was focused on a circular spot of about 2 mm in diameter. With the exception of XRD, all the above listed analyses were performed in the central area of the part of the substrate immersed into the electrolyte, i.e. within the area of the laser spot, where oxygen evolution occurred under illumination.

XRD analysis yielded no significant differences (formation of secondary phases, changes of lattice parameter) before and after PEC. ERDA also yielded no changes of the nitrogen and oxygen contents in the bulk of the films, when comparing the samples before and after PEC. Only at the surface, within the limits of surface sensitivity of ERDA, for both samples (before and after PEC) a slight increased oxygen content and decreased nitrogen content could be observed.

LEIS depth profiling was performed to investigate the near-surface distribution of cations. The cation peak areas, proportional to the surface coverage, showed that the outer surface is La-rich compared to the bulk composition for both films, before and after PEC. From the depth profiles a comparison of the ratios between the Ti and La contents could be obtained, as shown in **Figure 5**a. Both films showed a similar near-surface Ti-depleted region that extends to about 1.5 nm. It should be mentioned that the extent of the Ti-depleted region might be limited to the first or second atomic layers, even though the depth profile reveals a deeper length. This is explained with the ionic mixing produced by the 500 eV Ar$^+$ sputtering (the projected range of the sputter ions is around 1 nm as estimated for other perovs.kite materials).

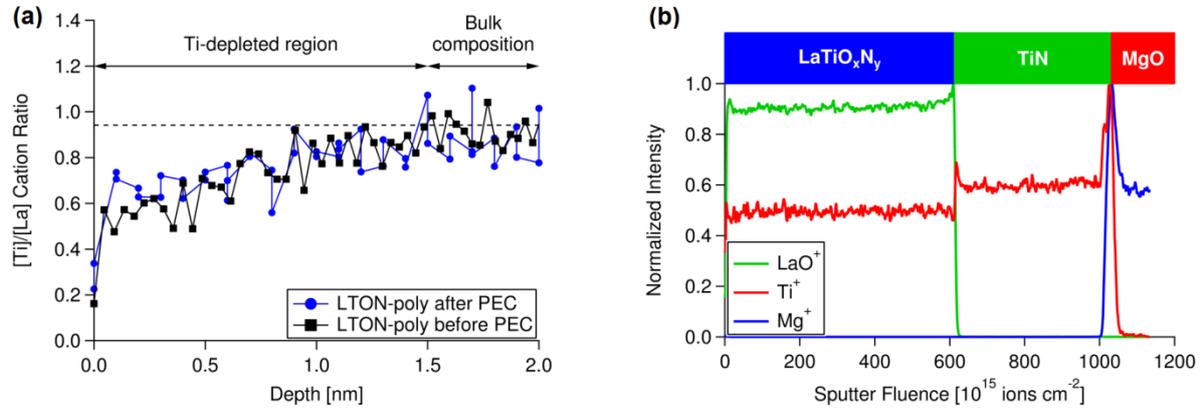

**Figure 5.** (a) Ti/La-ratio of LTON-poly before and after PEC normalized to the bulk stoichiometry as obtained from the LEIS depth profiles. (b) Depth profile of LTON-poly before PEC using ToF-SIMS.

Such an increased lanthanum content (titanium depletion) on the surface of our films is in agreement with ref. [7a], where a reduction of the Ti/La surface atomic ratio for LaTiO$_2$N nanoparticles was observed after nitridation (Ti/La = 0.7), while the bulk remained unaffected (Ti/La = 1). However, literature also reports a Ti-enriched amorphous surface for LaTiO$_2$N powder samples after nitridation.[28] After annealing and chemical etching in aqua regia this Ti-rich amorphous layer was removed and the nanoparticles showed a reconstructed surface. ToF-SIMS chemical depth profiles showed remarkably well-defined interfaces between the different layers (TiN/MgO and LaTiO$_x$N$_y$/TiN) and a uniform chemical composition of each layer throughout the films in very good agreement with RBS and ERDA measurements (see Fig 5b). Incidentally, ToF-SIMS also indicated that the TiN buffer layer is not oxygen-free, as observed in previous studies[28] where it has been proven that this does not affect the PEC measurements. Interestingly, ToF-SIMS revealed that the nitrogen-to-oxygen ratio of the TiN layer is smaller at the MgO side suggesting diffusion of oxygen from the substrate into the TiN layer during deposition. This effect of oxygen diffusion into the deposited thin film is well known in literature for substrates like SrTiO$_3$ and LaAlO$_3$.[38]

ToF-SIMS also detected traces of surface contamination in the sample after PEC characterization not present in the bulk or at the surface of the sample before PEC. These contaminants could be assigned to chromium, manganese and iron. Also hydrocarbon-derived contaminants were found on the PEC tested samples. We assume that these contaminants originate from the electrolyte.

In summary, the structural and chemical composition of the samples did not show any significant difference as a consequence of the PEC measurements highlighting the good physicochemical stability of the LaTiO$_x$N$_y$ thin films. These findings also suggest that the reason for the initial reduction of the measured photocurrent has to be ascribed to changes of the local surface chemical environment of the constituent elements, rather than to changes in the chemical composition profile (i.e. loss of nitrogen).

X-ray Photoelectron Spectroscopy (XPS) was applied to gain further insights on this matter.[39] The XPS results for Ti 2p, N 1s, O 1s, and La 3d core levels are shown in **Figure 6**. The LaTiO$_x$N$_y$ sample before and after PEC were measured at normal incidence angle (90°), while the sample before PEC was also measured at 45° incidence angle to discriminate the components located at the surface.

As a first observation, comparing the XPS measurements before and after PEC (at normal incidence angle), no significant changes of the Ti/La, N/La and N/Ti atomic ratios could be detected, suggesting that the surface composition and therefore also the nitrogen content remain unaffected by the PEC measurements (no nitrogen loss).

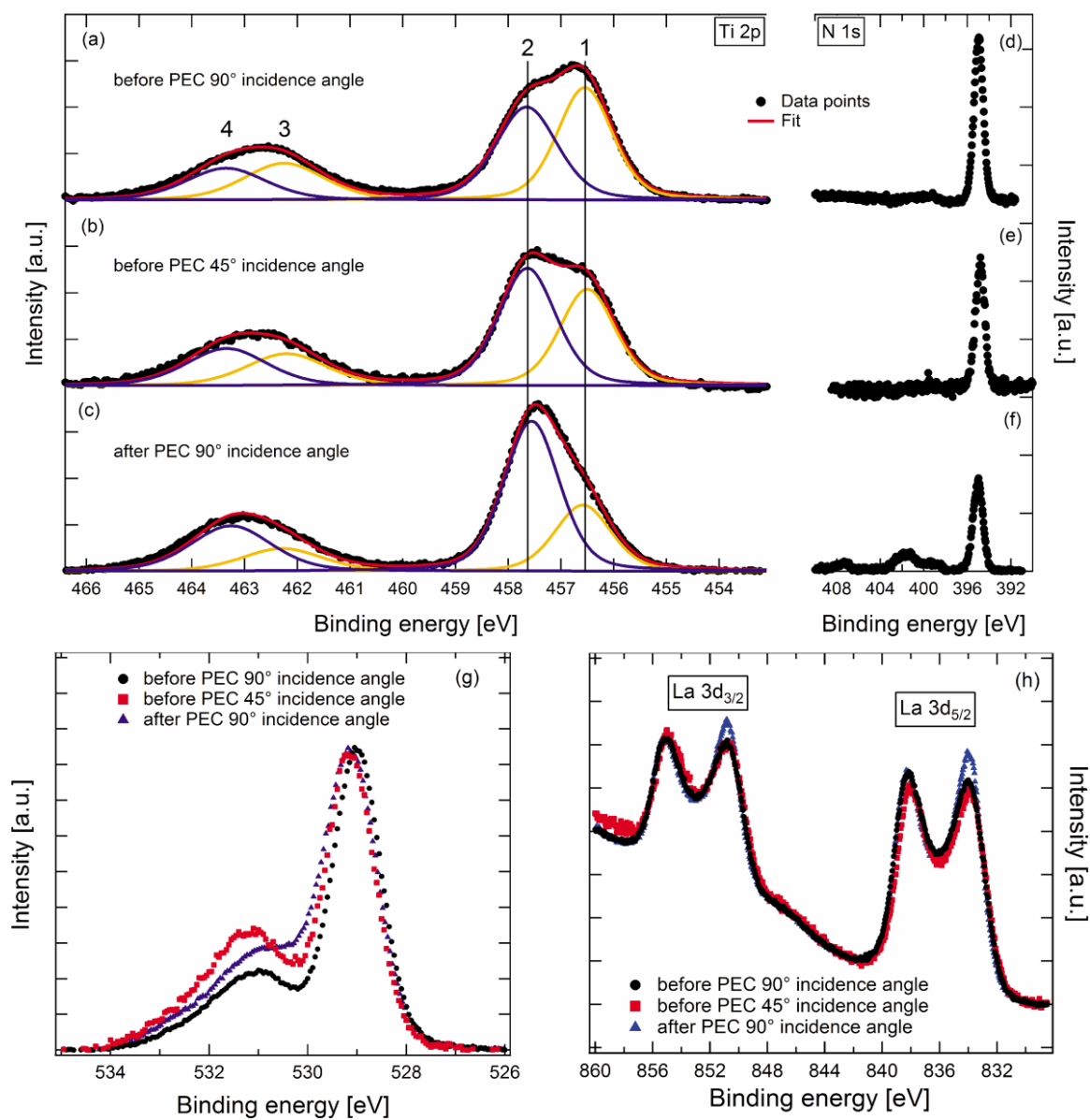

**Figure 6.** XPS spectra of (a)-(c) Ti 2p, and (d)-(f) N 1s, O 1s (g), and La 3d (h) of LTON-poly before PEC (incidence angle of 90° and 45°) and after PEC (incidence angle of 90°).

Concerning the nitrogen content comparing the samples before PEC at 90° and 45° incidence angle, smaller nitrogen content can be observed for the latter. This is in agreement with the increased oxygen content due to the LaO-terminated surface observed by LEIS (see Figure 5a).

An increased lanthanum surface content can be observed for both samples before and after PEC compared to bulk. This also confirms the preferential LaO surface termination (and Ti depletion) observed by LEIS (see Figure 5a). It is worth mentioning here that the XPS analyses provide quantitative values mediated over a surface layer of about 8 nm thick, while LEIS can assign this La-rich layer to the first nanometer, thus this LaO-termination is limited to the first or second atomic layer as mentioned above.

The analysis of the shifts in the Ti binding energies finally shows remarkable differences between the two samples before and after PEC.

Figure 6a-c shows the Ti 2p XPS spectra of the LaTiO$_x$N$_y$ films before PEC at 90° (Figure 6a) and 45° incidence angle (Figure 6b) as well as after PEC (Figure 6c). Two main peaks can be identified at a binding energy of about 457 eV and 463 eV and they can be assigned to Ti 2p$_{3/2}$ and Ti 2p$_{1/2}$ spin orbit, respectively. Each main peak can be fitted with two Gaussians, component 1 (orange) and 2 (blue) in Figure 6a-c. The peak positions of the two components are roughly the same for the three different measurements and are marked with two vertical lines in Figure 6a-c.

The main peak located at 457 eV suggests that titanium is present in both, the Ti$^{4+}$ oxidation state (Ti 2p peak for TiO$_2$ is ≈458 eV[40]) and in the Ti$^{3+}$ oxidation state (Ti 2p peak for TiN is ≈455 eV[40b-d, 41]). This was expected due to the under-stoichiometric nitrogen content of our LaTiO$_x$N$_y$ samples, which requires a mixture of Ti$^{4+}$ and Ti$^{3+}$ oxidation states, while in LaTiO$_2$N only Ti$^{4+}$ is present. Also for TiO$_x$N$_y$ the Ti 2p$_{3/2}$ peak was found between the Ti 2p peak positions of TiO$_2$ and TiN.[8, 40b-d], On the basis of these considerations, we ascribe the binding energy of the component 1 in Figure 6a-c to a less oxidized chemical environment (higher nitrogen content). Accordingly, the component 2 is assigned to a titanium chemical environment containing more oxygen (lower nitrogen content).

By comparing the Ti 2p XPS measurement at 90° and 45° incidence angle before PEC characterization in Figure 6a and Figure 6b, the binding energy of component 2 (more oxidized chemical environment) at 457.6 eV can be assigned to the titanium located closer to the LaO-terminated surface, since the peak intensity increases at 45° incidence angle compared to the measurement at normal incidence. Accordingly the binding energy of component 1 (less oxidized chemical environment) at about 456.6 eV can be assigned to the titanium mostly in the bulk. Comparing now Figure 6a-b with Figure 6c, we can see that the intensity of component 2 further increases after PEC characterization even though the XPS measurement was performed at normal incidence. This suggests that after PEC measurements titanium is more oxidized at the surface compared to the bulk.

Concerning the N 1s XPS spectra (Figure 6d-f), no significant changes could be observed at 90° (Figure 6d) and 45° (Figure 6e) incidence angle before PEC. The peak at the binding energy of about 395 eV is thus assigned to nitrogen in the chemical environment of $LaTiO_xN_y$. On the contrary, the comparison of the N 1s XPS spectra before PEC (Figure 6d) and after PEC (Figure 6f) exhibits clear changes. After PEC additional peaks, which are difficult to assign, appear at higher binding energies. In ref. [8] an additional peak of the N 1s XPS spectra at very similar binding energy was found after thermal annealing of $TiN_xO_y$ in air. This peak was assigned to N-C bonds, which were formed with carbon surface contamination. A very similar evolution of the N 1s spectrum was observed in a study of the titanium nitride oxidation chemistry.[42] In this work the additional N 1s spectra features appearing after oxidation were assigned to different types of bonding configuration for chemisorbed nitrogen.

According to the NIST X-ray Photoelectron Spectroscopy Database[43] for N 1s, the peaks that appear at higher binding energies in the N 1s spectrum (Figure 6f) can be assigned to nitrogen bonded with hydrogen, carbon and oxygen.

The broad N-1s peak measured after PEC in the region of 399-403 eV was also observed in ref. [17] analyzing $TiO_{2-x}N_x$ thin films fabricated by sputtering using a mixture of Ar, $O_2$ and $N_2$ gas. This peak was present also for $TiO_2$ films when the N 1s sharp peak at about 395 eV was, as expected, not detected (though, trace of $N_2$ was present in the $Ar/O_2$ gas mixture). In other studies this broad peak was assigned to adsorbed[42, 44] or interstitial[45] molecular $N_2$.

Since the XPS results shown here indicate that the overall nitrogen content of our films remains almost unchanged after PEC measurements, in analogy to literature suggestions we speculate that the change of the N 1s XPS spectrum is due to nitrogen atoms that leave the $LaTiO_xN_y$ structure, are replaced by oxygen and migrate to the surface to form adsorbed or interstitial molecular $N_2$ and/or to bond with hydrocarbon-derived contaminants as observed by ToF-SIMS. Literature reports that the nitrogen loss depends on the pH of the electrolyte with the effect becoming more important reducing the pH from 9 down to 4.[7a] To our best knowledge, no nitrogen loss was observed with a high pH ($\geq 13$).[37b] For this work an electrolyte with pH 13 was used thus making the nitrogen loss negligible.

The observed oxidation of titanium and the suggested replacement of nitrogen with oxygen at the surface should also lead to changes in the O 1s XPS spectra (see Figure 6g). However, the observed changes in the O 1s XPS spectra at higher binding energies may also be related to the adsorbed species at the sample surface, as revealed by ToF-SIMS. Whether or not the oxygen content is increased within the 8 nm of the XPS penetration depth cannot be identified and further investigations are required.

The discussed changes in Ti 2p and N 1s XPS spectra also lead to changes in the chemical environment of lanthanum. Figure 6h shows the La 3d multiplet, where the peaks at smaller binding energies of both doublets show increased intensities after PEC.

To summarize, the reported characterizations suggest that the observed initial degradation of the PEC performance of our $LaTiO_xN_y$ films may be due to the presence of a La-rich $LaTiO_xN_y$ layer where titanium undergoes further oxidation and nitrogen is partially depleted, in lattice position, during operation.

## 3.4. Effect of the Different Crystallographic Surface Orientations on the Visible Light Induced Photocurrent

Literature reports the comparison of the photo-response of different crystallographic surface orientations of TiO$_2$.[46] It was found for example that (001), (101) and (111)-oriented TiO$_2$ thin films prepared by sputtering showed the highest photoactivity toward the reduction of Ag$^+$ to Ag.[46a] By comparing the effect of photocatalytic reactions at differently oriented facets of rutile and anatase particles it was also found that certain crystallographic orientations provides the effective reduction or oxidation sites (for example the {110} and {011} planes of rutile, respectively).[46d]

To our knowledge, a comparative study of the photoelectrochemical response of different crystallographic surface orientations of oxynitride photocatalysts is addressed here for the first time. Yet, before tackling this aspect the above discussed initial evolution of the surface of the sample under investigation must be taken into account.

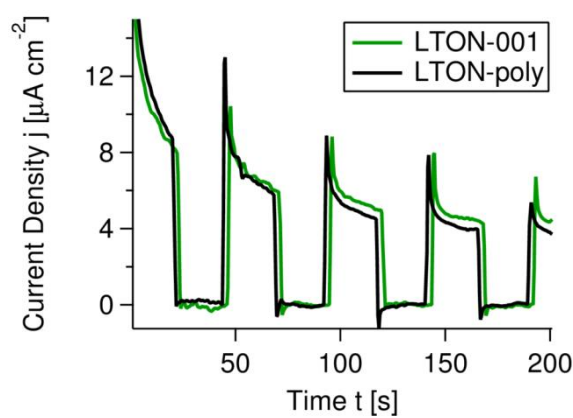

**Figure 7.** Potentiostatic measurements at 1.5 V vs. RHE of sample LTON-001 and LTON-poly (24 s light on and 24 s light off). All measurements were performed in an electrolyte of 0.5 M NaOH (pH = 13) using a 405 nm laser diode with a light intensity of 130 mW cm$^{-2}$. For the ease of comparison, the dark current is subtracted from the data.

As shown above, the PEC measurements performed using LTON powder samples and LTON polycrystalline films showed that the measured photocurrents gradually approach constant and reproducible values after few voltage scans (4 to 6, in our experimental conditions) probably as a result of the initial surface characteristics and evolution during operation. Moreover, PEC characterizations performed after sample stabilization by holding the potential stable (potentiostatic or chronoamperometric measurements) show that the photocurrent further decreases within the first few minutes before stabilizing. As an example, **Figure 7** shows the potentiostatic measurements of LTON-poly and LTON-001 performed by holding the applied voltage at 1.5 V vs. RHE for 200 s. Besides the initial spikes due to a charging and discharging of surface states or oxidation and reduction of surface species,[47] the photocurrents decrease from their initial values of more than 10 µA cm$^{-2}$, similar to those measured by potentiodynamic measurements at the same voltage after sample stabilization, down to about 6 µA cm$^{-2}$ and 4 µA cm$^{-2}$ for LTON-001 and LTON-poly, respectively. The extent of such a decrease of photocurrent and the time required for stabilization depends on the applied voltage and sample history (previously applied voltage, holding time, positive or negative voltage variation between two consecutive measurements). This effect may depend on adsorbed oxygen blocking the active sites, which hinder the electrochemical reaction. The use of appropriate co-catalysts may further improve the performance, but it would drive the present investigation toward the study of the properties of different electrochemical systems.

The purpose of this investigation is the comparison of the electrochemical properties of photoanode surfaces with different crystallographic orientations and the establishment of an experimental approach that will enable the extension of such a comparison to different materials. Previous considerations suggest that the typical anodic (from low to high voltage scan) potentiodymanic characterization widely used in the literature may be not the most reliable approach. Following such an approach in fact, for different surface orientations or using different materials would raise the questions: How many potentiodynamic scans are required to achieve full surface stabilization? How a potentiodynamic measurement is influenced by the voltage history of the sample under investigation? How much these effects of surface evolution differ for different crystallographic orientations? In addition, since the measured photocurrent is intrinsically small (in the range of few μA cm$^{-2}$ due to the much smaller electrochemically active area for thin films compared to powder samples), how much does the capacitive current, which is proportional to the voltage scan rate in a potentiondynamic measurement, affect the measured photocurrent?

These uncertainties have led us to select a different investigation approach, that is the comparison between potentiostatic photocurrent measurements performed during cathodic voltage scans (from high to low applied potential).

**Figure 8**a shows the results of these measurements. For each sample 1.5 V vs. RHE was first applied for 200 s, which reveals that the photocurrent decays, as shown in Figure 7. The voltage was then reduced in steps of 25 mV holding the potential constant for 200 s and the steady state photocurrent values at the different potentials are summarized in Figure 8a. After the first 200 s at the highest voltage, at all lower voltages the photocurrent stabilizes in much shorter times. Below 1.4 V vs. RHE only few seconds are needed to achieve a steady photocurrent density. The photocurrent measured at the highest voltage of 1.5 V vs. RHE was used as sample pre-conditioning but not taken into account in Figure 8a since, as shown in Figure 7, the photocurrent seems to be not yet fully stabilized after 200 s. The difference between the anodic potentiodynamic measurement and the cathodic potentiostatic measurement of LTON-001 is shown in Figure 8b as a representative example. The cathodic potentiostatic photocurrent is significantly smaller (a factor of 5 at 1.4 V vs. RHE, as an example) but we do believe that this measurement is much less sensitive to spurious effects and/or sample history.

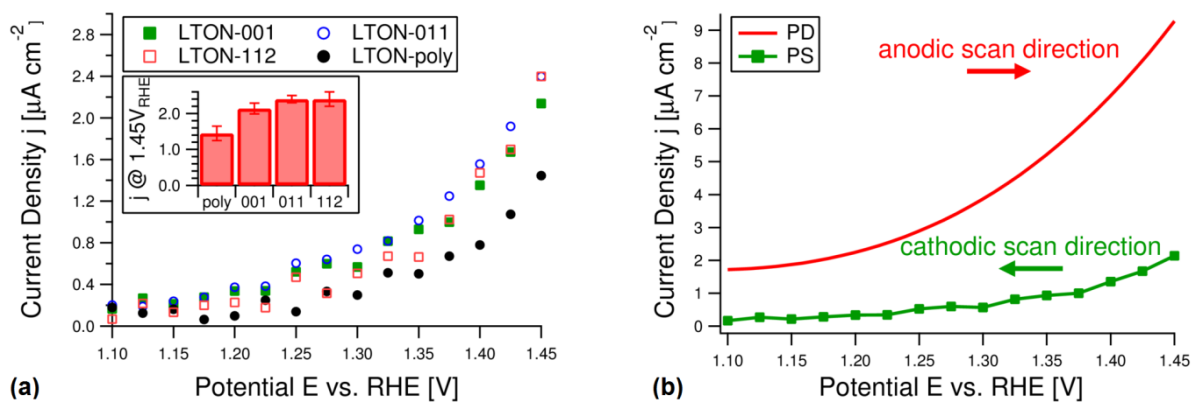

**Figure 8.** (a) PEC characterization of the (001)-oriented sample LTON-001 (open green squares), the (011)-oriented sample LTON-011 (open blue circles), the (112)-oriented sample LTON-112 (full red squares), the polycrystalline sample LTON-poly (full black circles). Every experimental value in this j-E curve was obtained from a single potentiostatic measurement using the value of the stabilized and dark current subtracted photocurrent. (b) the 4$^{th}$ potentiodynamic measurement in comparison with potentiostatic measurement using sample LTON-001. The scan direction for the potentiodynamic measurement is from low to high voltage (anodic scan) and the scan direction for the potentiostatic measurement is from high to low voltage (cathodic scan).

Figure 8a clearly shows that, in agreement with previous studies[19], crystalline quality matters. As an example, at 1.45 V vs. RHE about 60% larger photocurrents were measured with the textured and epitaxial samples.

Figure 8a also shows that similar photocurrent densities were measured for the three different orientations. However, the different nitrogen content and, as a consequence, the different light absorption properties must be taken into account.

As previously pointed out, the nitrogen content of the LaTiO$_x$N$_y$ films increases with increasing the laser fluence at the expense of the crystalline quality of the films, as shown in Table 2. Epitaxial films have in fact lower nitrogen-to-oxygen ratios (in the range of 0.13 to 0.20) than polycrystalline or textured films (about 0.4). By comparison with the UV-Vis measurements in Figure 3a, in our experimental condition we can estimate (neglecting reflectance) a light absorption of about 95% for LTON-poly and LTON-112, 50% for LTON-011 and 30% for LTON-001. The light absorbed in the underlying TiN layer does not affect the photocurrent as confirmed in ref. [28] and shown in Figure S1.

A direct measurement of the transmittance for the same samples used for PEC characterization is not possible due to the presence of the TiN layer. To estimate the absorption of the different samples thin films grown with N/O ratio of 0, 0.18, 0.20, 0.27, and 0.33 on transparent double-side polished substrates were used (see Figure 3a and reference 28). For N/O ratio around 0.33 the absorption at 405 nm is already well above 90%. A value of 95% for LTON-poly and LTON-112 was estimated considering the broadening of the absorption spectrum observed by increasing the N content. LTON-110 shows N/O ratio of about 0.19 and can be directly compared to the measurements performed on samples with N/O ratio of 0.18 and 0.20. For LTON-001 and absorption value of about 30% was estimated considering the measurements performed on samples with N/O ratio of 0.20, 0.18, and 0 and assuming an almost linear dependence between absorption and N content. This assumption is reasonable for small N/O ratio where the effect of broadening of the transmittance spectra is negligible.

In view of these considerations, the PEC performance of the 4 samples can be compared on the base of the absorbed photons-to-current conversion efficiency (APCE) defined as

$$APCE\ (\lambda) = \frac{hc}{e} \frac{J_{ph}(\lambda)}{\lambda A P_{in}(\lambda)}$$

where $J_{ph}(\lambda)$ is the measured current density at a specific applied potential, $P_{in}(\lambda)$ is the power density of the monochromatic light source at the semiconductor surface, A is the light absorption (in percent), h the Planck constant, c the speed of light and e the elementary charge (hc/e = 1240 m$^2$ g$^{-1}$). We conclude that while the measured photocurrent densities of the three crystallographically ordered samples were similar, the respective APCE is indeed significantly different. As an example, **Table 3** shows the APCE values of the four samples evaluated at an applied potential of 1.45 V vs. RHE.

**Table 3.** Absorbed-photons-to-current efficiency values of the four described samples with $P_{in}$=130 mW cm$^{-2}$, $\lambda$=405 nm and the obtained photocurrent densities ($J_{ph}(\lambda)$) from Figure 8a at an applied potential of 1.45 V vs. RHE.

| Sample | Absorption | Photocurrent density | APCE |
|---|---|---|---|
| | | µA cm$^{-2}$ at 1.45 V vs. RHE | 1e-5 at 1.45 V vs. RHE |
| LTON-poly | 95% | 1.455 | 3.6 |
| LTON-112 | 95% | 2.399 | 5.9 |
| LTON-011 | 50% | 2.398 | 11.3 |
| LTON-001 | 30% | 2.138 | 16.8 |

It is interesting to note that the polycrystalline (LTON-poly) and the textured (LTON-112) samples are both mainly (112)-oriented and have very similar nitrogen contents and optical properties. By comparing the PEC measurements of these two samples, we observe the effect of the different crystalline quality. The film with higher crystalline quality shows an increase in PEC performance of about 65%. Instead, for the two epitaxial films (LTON-011 and LTON-001) the effect of high crystalline quality is combined with the influence of the different surface orientation. The (001)-oriented surface shows a value of APCE about 50% larger than that obtained for the (011)-oriented surface. The (001)-oriented samples show an APCE value about 5 times higher than that of the polycrystalline sample.

It is worth highlighting here that due to the specific optical properties of the samples, the use of a different visible-light source instead of a laser diode with wavelength of 405 nm does not significantly affect the results. The two samples with N/O ratio below 0.20 are already largely transparent at this wavelength. Concerning the two samples with N/O ratio around 0.40, almost the same photocurrent was measured using the laser diode and a Xe arc lamp, as shown in Supporting Information, Figure S3.

We conclude that the crystalline quality of the semiconductor certainly affects the PEC performance by reducing the number of recombination sites. In addition, we observed that the crystallographic surface orientation plays also an important role, most probably by influencing the charge transfer at the semiconductor/electrolyte interface.

To probe this hypothesis we performed density functional theory (DFT) calculations to predict the structure of the surfaces (in terms of atomic arrangement of the different crystallographic planes) and to gain more insights into the thermodynamic stability of the different surfaces.

To compare with the LTON-(001) sample, the LaO (**Figure 9**a), the LaN (Figure 9b), and $TiO_2$ (Figure 9c) surface terminations of the (001) surface were calculated. We compute a surface energy of 0.542 J m$^{-2}$ of the LaN termination (Figure 9b), which has the lowest overall surface energy with the $TiO_2$ termination (Figure 9c) being the second most stable surface at 0.654 J m$^{-2}$. The LaO termination (Figure 9a) is quite unstable with a surface energy of 0.980 J m$^{-2}$, and is expected to occur less frequently than the LaN termination. Surface energies for other cation orderings are generally higher and will be reported elsewhere.[48] This energetic preference for a La terminated (001) surface agrees with the presented LEIS data. Concerning the mostly (112)-oriented samples, the most stable termination of the (112) surface (0.875 J m$^{-2}$, Figure 9d) contain mixed cation layers with a 1:1 ratio between La and Ti and is hence unlikely to contribute to a La concentration enhancement unless antisite defects or Ti vacancies are formed. This implies that the polycrystalline sample should contain a fraction of (001) surfaces or other reconstructions that enhance the La concentration at the surface.

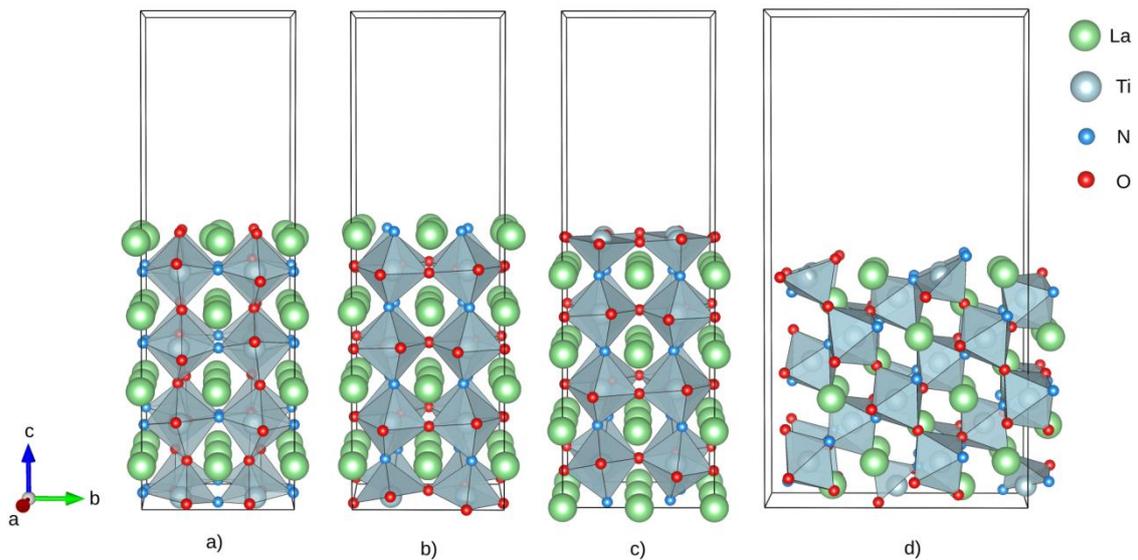

**Figure 9.** a) LaO (0.980 J/m$^2$), b) LaN (0.542 J/m$^2$) and c) TiO$_2$ (0.654 J/m$^2$) terminations of the (001) surface for the energetically most preferential anion order as well as d) energetically most preferential structure of the (112) surface (0.875 J/m$^2$).

The low surface energy of the (001) LaN termination compared to the LaO termination is a measure of the high resistance of the (001) surface towards oxidation. The (112) surface on the other hand prefers a mixed N/O termination and is likely to be more easily oxidized. Besides this auto-passivation, the higher PEC activity of the (001) surface can also be related to the local atomic composition of the energetically most favorable LaN termination. In **Figure 10**a we show the layer-resolved electronic densities of states (DOS) for the first and second layer of the LaN terminated (001) surface, whereas in Figure 10b we show the ones for the LaO terminated surface. The dashed vertical lines indicate the highest valence band states in the respective layer. We can see that N-containing layers have valence-band states at higher energies than O-containing layers, which stems from the higher electronegativity of O compared to N. This implies that a hole migrating from the subsurface to the surface, where it can oxidize water, will move to a more favorable (higher energy) state on the LaN termination, whereas on the LaO termination this hole migration is endothermic. The energy differences are of the order of 0.1 eV and the presence of N on the surface is therefore expected to positively affect the hole-kinetics in the surface region.

We thus argue that the PEC activity of a photoanode, which is determined by the ease of hole transfer to water molecules on the surface, depends on the concentration of N in the surface layer. For the (001) surface, we compute the lowest energy for a surface containing only N anions (Figure 9b), whereas for the (112) surface (Figure 9d), the ratio between electrolyte-accessible N and O is 3:7 for the energetically most favorable surface termination. Based on these results, the (001) surface should have a higher PEC activity than the (112) surface, which agrees with the (001) surface having a significantly higher APCE. The (011) surface is strongly polar and is expected to reconstruct, we do however at present not have a clear picture of its structure and N content.

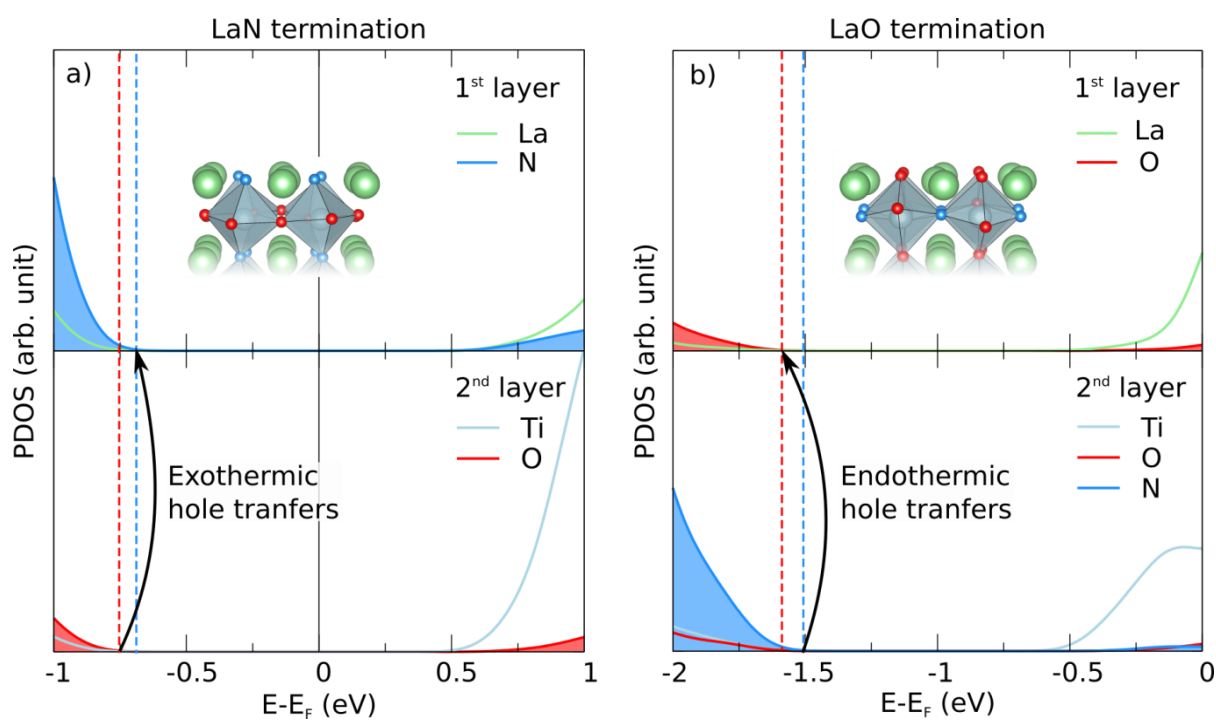

**Figure 10.** Layer-resolved electronic densities of states for the first and second layer of a) the LaN terminated (001) surface and b) the LaO terminated (001) surface. Dashed vertical lines indicate the top of the valence band in the respective layers and arrows show the migration of holes from the subsurface to the surface layer. Note that the LaO termination compensates polarity by self-doping with electrons, resulting in a Fermi energy in the conduction band, which does however not affect the conclusions drawn here.

The availability of different crystallographic surface orientations will allow investigating the role of suitable co-catalysts with different interfaces with the semiconducting photoanode. This is of course essential to effectively make the photo-generated charge carriers available for the electrochemical reaction.

## 4. Conclusion

We investigated the effect of the crystalline quality and crystallographic surface orientation of $LaTiO_xN_y$ thin films on the visible light induced photocurrent using TiN-buffered substrates as current collectors.

The analysis of the chemical composition of the as prepared sample surfaces revealed the presence of a La-terminated surface layer, where during the first electrochemical tests titanium undergoes further oxidation and nitrogen is partially depleted in the lattice sites. No significant nitrogen loss was observed in the applied experimental conditions.

We showed that such chemical evolution of the surface reduces the measured current. A large part of the current measured during the first potentiodynamic tests can possibly be ascribed to this electrochemical surface oxidation and is not (or at least not only) the result of the water splitting process. The results of the first potentiodynamic measurements are largely related to the specific sample fabrication and experimental setup, and should not be used to compare the PEC performance of different samples and/or materials. After few potentiodynamic scans the current reaches a stable and reproducible value which we assign to the real photocurrent associated to visible light induced water splitting.

In order to characterize thin film photoanodes with different crystallographic properties after chemical stabilization and minimize the influence of spurious effects and/or voltage history, we propose the use of potentiostatic cathodic voltage scans as alternative to the potentiodynamic anodic measurements typically applied in literature.

Our measurements confirm that the crystalline quality of the sample has an important influence on the PEC performance; better crystalline quality increases the attainable photocurrent for semiconducting photoanode with comparable optical properties. Moreover, we also show that the absorbed photon-to-current conversion efficiency can be tuned by selecting different crystallographic surface orientations. In particular, the (001)-oriented surface shows an absorbed photon-to-current conversion efficiency almost three times higher than the (112)-oriented surface, which is the XRD reflex with the largest relative intensity for this material, thus possibly the energetically favored surface orientation for $LaTiO_xN_y$ nanoparticles. Furthermore, the absorbed-photon-to-current conversion efficiency of (001)-oriented films is about five times higher than for polycrystalline samples.

Density functional theory calculations support the experimental findings and suggest that the PEC activity of a photoanode, which is determined (among other parameters) by the ease of hole transfer to water molecules on the surface, depends on the concentration of N in the surface layer. Layer-resolved density of state calculations show in fact that N-containing surface layers (e.g. a LaN termination of the (001)-oriented surface) have valence-band states at higher energies than O-containing layers. Such a local surface chemical environment and energetic landscape favours the migration of the photo-generated holes towards the solid-liquid interface.

The results of this investigation point at new goals to achieve next: optimize the chemical composition of $LaTiO_xN_y$ films (bulk and surface), increase the nitrogen content of highly ordered films, and match selected surface orientations with suitable co-catalysts.

## 5. Experimental Section

*Thin film deposition:* LaTiO$_x$N$_y$ thin films were grown using a modified Pulsed Laser Deposition (PLD) method called Pulsed Reactive Crossed-beam Laser Ablation (PRCLA). PRCLA uses a piezoelectric nozzle valve to inject pulsed gas jets near the ablation spot at the target surface. Using the laser as the trigger for the piezoelectric actuator, the appropriate opening time and delay of the opening time with respect to the laser pulse are selected. The reactive gas jets intersect the ablation plume close to the ablation spot, i.e. at the very early stage of the plasma propagation, setting a strong pressure gradient of the reactive gas along the direction of the expansion of the ablation plume. This allows efficient physicochemical interactions of the plasma species with the gas jets near the target (high pressure) and almost collision free propagation of the plasma near the substrate (low pressure), hindering further physicochemical interactions of the plasma species formed at the early stage of the plasma propagation. More details about this modified PLD technique can be found in ref. [27-29].

In this study, a KrF excimer laser ($\lambda$=248 nm, pulse width of 30 ns) was used to ablate a sintered ceramic target of La$_2$Ti$_2$O$_7$ fabricated in our laboratory. A gas inlet line allowed to set a N$_2$ partial pressure of $8.0\times10^{-4}$ mbar in the vacuum chamber, while NH$_3$ gas jets were injected through the nozzle valve. The opening time of the nozzle valve was 400 µs and the delay between the opening time and the laser was 30 µs. The laser, and as a consequence the nozzle valve, was operated at a repetition rate of 10 Hz. In these conditions, the average pressure in the chamber raised to $1.8\times10^{-3}$ mbar. The laser fluences used for LaTiO$_x$N$_y$ thin films were in the range of 2.2-3.8 J cm$^{-2}$. The area of the ablation spot was 1 mm$^2$ and the target-to-substrate distance was 50 mm.

TiN layer needed as a current collector for photoelectrochemical measurements of LaTiO$_x$N$_y$ thin films was grown in situ on the selected substrates *via* conventional PLD in vacuum (base pressure of $5 \times 10^{-6}$ mbar) using a commercially available TiN target. A laser fluence of about 3.5 J cm$^{-2}$ and a laser repetition rate of 10 Hz were applied.

The (001)-oriented MgO and (0001)-oriented Al$_2$O$_3$ single crystal substrates were ultrasonically cleaned with ethanol and acetone. The deposition temperature was set to 800 °C for both LaTiO$_x$N$_y$ and TiN. Pt paste was used to provide good thermal contact between the substrate and the heating stage. The substrate temperature was monitored with a pyrometer.

*LaTiO$_2$N Powder and Photoanode Preparation:* LaTiO$_2$N (LTON) particles were prepared *via* thermal ammonolysis of La$_2$Ti$_2$O$_7$ at 950 °C for 11 h under an ammonia flow of 250 mL min$^{-1}$. La$_2$Ti$_2$O$_7$ precursor was obtained by the solid-state route whereby a stoichiometric mixture of La$_2$O$_3$ and TiO$_2$ was calcined at 1150 °C for 5 h in the presence of NaCl flux. LTON photoanodes were fabricated using electrophoretic deposition (EPD).[49] In details, two fluorine-doped tin oxide (FTO) substrates ($1 \times 2$ cm$^2$) with a distance of 1 cm were immersed in a suspension of LTON particles with their conducting sides facing each other. The suspension was prepared by mixing LTON powders (60 mg) and iodine (15 mg) in acetone (75 mL). A bias of 20 V was applied for 10 min in which the positively charged particles were deposited at the negative electrode. A post-necking treatment was applied to enhance the contact between LTON particles as described elsewhere in literature.[36, 49] The electrodes were dropped with TaCl$_5$ (30 µL) methanol solution (10 mM) and then dried. After repeating this process for 3 times, the electrodes were heated in air at 300 °C for 30 min followed by annealing in ammonia at 400 °C for 1 h.

*Photoelectrochemical characterization:* Photoelectrochemical (PEC) measurements were performed using LaTiO$_x$N$_y$ thin films grown on TiN-buffered MgO and Al$_2$O$_3$ substrates in a three-electrode configuration. LaTiO$_x$N$_y$ thin films and Pt wire were the working electrode and the counter electrode, respectively. The reference electrode was a Ag/AgCl electrode immersed in a saturated KCl electrolyte.

TiN is not transparent, thus only front-side illumination was possible. An aqueous solution of 0.5 M NaOH (pH=13) was used as electrolyte. At one edge of the sample, the $LaTiO_xN_y$ layer was carefully removed to expose the TiN layer for the electrical contact to the read-out of the potentiostat (Solartron 1286 electrochemical interface). The samples were rinsed with de-ionized water and only the $LaTiO_xN_y$-coated part of the sample was immersed into the electrolyte. The samples were illuminated with a 405 nm laser diode (Laser 2000) with 5 mW power output and a spot size of about 0.03 cm$^2$. The light intensity was estimated to be about 130 mW cm$^{-2}$ considering reflection losses at the quartz cell and at the surface of the sample. To measure the dark current (no light on the sample) and the photocurrent consecutively, an asymmetrical time lag relay pulse generator was used to intermittently irradiate the sample.

For potentiodynamic measurements, we chose a scan rate of 10 mV s$^{-1}$ and a potential range between 0.1 V and 1.5 V vs. RHE (anodic scan). Potentiostatic measurements were performed between 1.1 V and 1.5 V vs. RHE (cathodic scan) in potential steps of 25 mV holding the potential constant for 200 s.

The same cell configuration used for LTON thin films was used for the photoelectrochemical characterization of LTON powder samples. However, the front-side illumination of the LTON powder photoanodes was realized using a 150 W Xe arc lamp (Newport 66477) with an AM 1.5G filter with an output intensity of 100 mW cm$^{-2}$.

*Thin film characterizations:* To determine the bulk chemical composition, Rutherford Backscattering (RBS) measurements were performed. The samples were irradiated with a 2 MeV $^4$He beam and the backscattered $^4$He particles were detected with a silicon PIN diode detector under an angle of 168 ° towards the incident beam. Data collected by RBS were analyzed by the RUMP program.[50]

Heavy-Ion Elastic Recoil Detection Analysis (ERDA) was used to determine the nitrogen-to-oxygen ratio. Here a 13 MeV $^{127}$I beam was directed on the sample and a combination of a time-of-flight spectrometer with a gas ionization detector were used for the detection of the recoiled particles.[51]

A Bruker-Siemens D500 X-Ray Diffractometer with a characteristic Cu K$_α$ radiation (1.54 Å) was used to investigate the crystalline properties.

In the wavelength range between 200 nm and 1000 nm (6.20-1.24 eV) the transmittance spectra were measured with a Cary 500 Scan UV-Vis-NIR spectrophotometer.

A Zeiss Supra VP55 Scanning Electron Microscope with an in-lens detector enabled the investigation of the cross sectional morphology.

An XE100 Atomic Force Microscope from Park Systems was used to determine the surface topography of the samples.

X-ray Photoelectron Spectroscopy measurements were performed using a VG ESCALAB 220iXL spectrometer (Thermo Fischer Scientific) equipped with an Al K$_α$ monochromatic source and a magnetic lens system. The films named "Before PEC" were analyzed as grown, while the films named "After PEC" were rinsed with de-ionized water followed by N$_2$ blow dry to remove the electrolyte before loading into the vacuum chamber for analysis.

The binding energies of the acquired spectra were calibrated using the C 1s line at 284.6 eV. Background subtraction has been performed according to Shirley,[52] and the atomic sensitivity factors (ASF) of Scofield were applied to estimate the atomic composition.[53]

Time-of-Flight Secondary Ion Mass Spectrometry (ToF-SIMS) analyses were performed on a ToF-SIMS V instrument (IonToF GmbH., Germany), using a bunched 30 keV Bi$^+$ analytical beam. Depth profiling was performed by using a second ion beam of 1 keV Cs$^+$ at 45° incidence.

To compensate for charging of the samples under the high currents of the sputter beam, low energy electrons were supplied by a flood gun. The use of cesium to perform the sputtering affords the analysis of $Cs_nM^+$ clusters, which serves two purposes; firstly, it mitigates the difference in secondary ion yields as the profile proceeds through different matrices ($LaTiO_xN_y$/TiN/MgO), and secondly, $Cs_2N^+$ and $Cs_2O^+$ clusters can be monitored to follow the depth distributions of oxygen and nitrogen. The latter is beneficial in the present case, because of the complicated overlapping distribution of $TiO^+$ and $TiN^+$ fragment ions corresponding to the isotope distribution of titanium.

Surface and near-surface characterization of the polycrystalline $LaTiO_xN_y$ thin film was performed by Low-Energy Ion Scattering (LEIS) (Qtac100 spectrometer, Ion-ToF GmbH., Germany). LEIS provides information of the elemental chemical composition of the first atomic layer of the surface by analyzing the energy distribution of the backscattered noble gas ions.[54] After introduction of samples into the LEIS instrument, they are exposed to a neutral flux of reactive atomic oxygen at room temperature that removes any adventitious contamination that would hinder the detection of the elements present at the outer surface. This cleaning method avoids any surface damage or atomic rearrangements that could be induced by other standard cleaning procedures, such as ion sputtering or annealing at high temperature.[55]

The primary ion beam is produced by a high brightness plasma source and directed towards the surface at normal incidence, while the backscattered ions are collected at a scattering angle of 145° from the entire azimuth by a double toroidal analyzer. Different primary ion species were used for the characterization in order to obtain information of light elements (e.g. light elements with a mass <20 u are detected by 3 keV $He^+$ primary ions) and good sensitivity and resolution for the heavier cationic species (analysis by 6 keV $Ne^+$ primary ions). The analysis areas were chosen to be rather large (1 mm$^2$) in order to minimize the surface damage during the analysis by keeping a low ion dose, while assuring that random distribution of grain orientations in the polycrystalline LaTiO$_x$N$_y$ surface was probed. Nevertheless, some light species present at the outer surface (e.g. light species present at the surface or adsorbed molecules from the residual gas) might be sputtered during the analysis using 6 keV Ne$^+$ ions, leading to a background signal that makes the detection of light cations difficult. In order to improve the detection of Ti at the surface, the cation surface analysis was performed by time-of-flight (ToF) filtering of the species arriving to detector, allowing to discriminate the secondary ion signal from the signal originated from Ne$^+$ primary ions scattered by Ti atoms at the surface.[56]

In addition to the surface characterization, the near-surface distribution of the cations was investigated by LEIS depth profiling. In this case, a dual beam analysis was performed by alternating the analysis with the 6 keV Ne$^+$ primary ions and a sputter cycle using a second sputtering beam. The sputtering was performed using a low-energy Ar$^+$ source (500 eV) at an incidence angle of 59°. To avoid edge effects during depth profiling, the sputter Ar$^+$ beam was rastered over a 1.3 × 1.3 mm$^2$ area.

*Density Functional Theory (DFT) Simulations:* The DFT calculations were carried out using the Quantum ESPRESSO package,[57] using the PBE functional.[58] Ultrasoft pseudopotentials are used with a kinetic-energy cut-off of 40 Ry and a cutoff for the augmented density of 320 Ry. A Hubbard U of 3.0 eV is applied on the Ti 3d orbitals. The experimental structure of Clarke et al.[59] was used to construct the slabs. The surface-cell size was 7.88 x 7.88 Å and 5.59 x 13.62 Å for the (001) and (112) slabs respectively and vacuum width of at least 10 Å was used to separate periodic images along the surface normal direction. Monkhorst-Pack[60] k-points meshes of 6x6x1 and 8x2x1 are used to sample the Brillouin zone for the (001) and (112) surfaces, respectively.

During structural optimization to a force threshold of 0.03 eV Å$^{-1}$, the bottommost unit cell layer is kept fixed at bulk positions. The surface energy is calculated following the equation: $E_{surf}=(2E_{rlx}-E_{unrlx}-E_{bulk})/2A$, where $E_{rlx}$ is the energy of the relaxed slab, $E_{unrlx}$ is the energy of the unrelaxed slab and $E_{bulk}$ is the energy of the bulk.

**Supporting Information**

Supporting Information is available from the Wiley Online Library or from the author.

**Acknowledgements**

The authors would like to thank the Paul Scherrer Institut, the Swiss National Science Foundation (IZERZ0 142176), and the National Centre for Competence in Research Discovery of new Materials (MARVEL) for financial support. SN and UA were supported by the Swiss National Science Foundation Professorship Grant PP00P2_157615.